# The Medical Authority of AI: A Study of AI-enabled Consumer-facing Health Technology

Yue You, YY, and You

College of Information Sciences and Technology, The Penn State University, University Park, PA, USA, yxy340@psu.edu

Yubo Kou, YK, and Kou

College of Information Sciences and Technology, The Penn State University, University Park, PA, USA, yubokou@psu.edu

Xianghua Ding, XD, and Ding

School of Computer Science, Fudan University, Shanghai, China, xianghua.ding@gmail.com

Xinning Gui, XG, and Gui

College of Information Sciences and Technology, The Penn State University, University Park, PA, USA, xinninggui@psu.edu

Recently, consumer-facing health technologies such as Artificial Intelligence (AI)-based symptom checkers (AISCs) have sprung up in everyday healthcare practice. AISCs solicit symptom information from users and provide medical suggestions and possible diagnoses, a responsibility that people usually entrust with real-person authorities such as physicians and expert patients. Thus, the advent of AISCs begs a question of whether and how they transform the notion of medical authority in people's everyday healthcare practice. To answer this question, we conducted an interview study with thirty AISC users. We found that users assess the medical authority of AISCs using various factors including AISCs' automated decisions and interaction design patterns, associations with established medical authorities like hospitals, and comparisons with other health technologies. We reveal how AISCs are used in healthcare delivery, discuss how AI transforms conventional understandings of medical authority, and derive implications for designing AI-enabled health technology.

CCS CONCEPTS • Human-centered computing~Human computer interaction (HCI)~Empirical studies in HCI

## 1 INTRODUCTION

Healthcare consumers' perception of medical authority keeps evolving along with technological advances [47,93,124]. Different from the notion of trust, authority has a structural and sociotechnical property, distributed and embodied within organizational members [11]. In this paper, medical authority refers to the power distributed in the whole healthcare socio-technical context, involving all stakeholders such as traditional authorities and health technologies. Healthcare consumers' medical authority perception and experience play an important role in whom and what venues they deem trustworthy and seek information from. The Internet allows people to turn to other people including strangers and nonprofessionals for medical information [18,109], through which they ascribe medical authority to these online sources. Such ascription of authority also takes place as healthcare consumers seek potential diagnoses from Artificial Intelligence (AI)-based symptom checkers (AISCs, or AISC apps). AISC apps can assess users' symptoms and provide potential diagnostic results. AISC apps have been popular in the healthcare market, and were used more than 100 million times in 2019 in the United States alone [89].

Consumer-facing AI health technologies bring radical changes to our very idea of medical authority. Usually, when people seek assessments for their symptoms, they ask help from doctors or ask their symptoms online. During this process, people are inclined to delegate authority to doctors and online peers [44,93,127]. With the advent of AI-enabled health technologies, new dynamics have been introduced into the existing network of medical authority. Unlike AI technologies designed for clinical practice, whose accuracy and authority can be evaluated by established authorities (e.g., medical professionals [24]), the authority of consumer-facing AI health technologies usually cannot be assessed by their users since they are laypersons without sufficient medical knowledge. If users blindly trust the diagnoses from consumer-facing AI health technologies, they may risk their lives [40], since these systems may generate inaccurate diagnoses and cause unintended consequences [63]. In this situation, it is important to know how users perceive and understand medical authority with the emergence of consumer-facing AI health technologies.

The concern for the medical authority of AI is also bound up with multiple growing concerns about AI systems, such as transparency [35,36], accountability [74], and authority [34]. In healthcare, a stream of research has investigated the accuracy [57,111] as well as usability [84,87] of AISC apps, factors influencing the authority of medical professionals [47,59], and the effect of online social forums [39,133] and Internet searches [120] on medical authority. Yet, few studies examined the authority of AISC apps and how these apps influence people's perception and experience of medical authority. Also, without such knowledge, it is difficult for designers and researchers to improve the future design of AISC apps. It is in this context we ask how the AISC apps impact how their users understand and experience medical authority.

To answer this research question, we conducted 30 semi-structured interviews with AISC app users in China. Many popular medical apps in China, such as Chunyu Doctor, Zuoshou Doctor, and WeDoctor, have already embedded an AI-enabled symptom checking function. Hence, AISC apps in China are ideal for us to investigate how users' understanding of medical authority to acquire rich data and insights. We found that the medical authority was dynamically constructed in users' interactions with AISC apps. First, the medical authority of AISC apps was associated with the endorsements from established authorities, such as reputable companies and hospitals. Second, how the user experience of AISC apps is designed also influenced participants' authority perception. Lastly, participants actively compared the results of AISC apps with results from another healthcare entities or organizations to adjust their medical authority perception. Based on these findings, we discuss the sociotechnical nature of the medical authority in the area of AISC apps and reflect upon the relationship between medical authority and interaction design. Lastly, we provide implications for designing AISC apps.

Our contributions are multi-fold: First, we contribute a deeper understanding of the authority of AI situated in the intersection of health informatics and HCI. Second, we articulate and foreground the notion of medical authority in understanding AI-enabled health technologies. Third, we deliver empirical and conceptual insights into how users interact with AISC apps and how the use influences their clinic visits, a growing phenomenon that has been understudied. Fourth, we shed light on the development and interaction design of AI-powered healthcare systems.

## 2 LITERATURE REVIEW

In this section, we will discuss existing studies on AISC apps, authority in medicine, and authority in AI technology, as well as how our study expands these lines of inquiry.

### 2.1 AISC apps

An AI-enabled symptom checker, also known as a self-diagnosing tool, is a type of consumer-facing digital health tool using AI techniques. They provide potential diagnoses and assist with triage (e.g., inform users whether they should seek medical care) [112]. Most studies on AISC apps focus on evaluating their effectiveness. Several studies have evaluated AISC apps' diagnostic accuracy by engaging multiple medical experts to assess checkers' features and performance through clinical vignettes [57,111]. While Babylon Health Inc. claimed their symptom checker, Babylon, had accuracy comparable to medical experts [85,103], a large amount of research has found that AISC apps were less accurate than physicians. In 2015, the British Medical Journal evaluated twenty-three AISC apps, disclosing that symptom checkers have proposed the correct diagnoses only 34 percent of the time [112]. In addition, the diagnostic capabilities of AISC apps are inferior to the diagnostic capabilities of medical professionals for certain diseases and conditions, such as HIV or hepatitis C [12], some ophthalmic conditions [115], DCM symptoms [31], and inflammatory arthritis [101].

Several studies have investigated the usability of AISC apps [84,87]. They found that users' evaluation of AISC apps changed in interactions [102]. One study also found the accuracy of diagnostic results depended on effective interactions between symptom checkers and users [77]. The researchers in this study found when users interacted with symptom checkers, they encountered problems regarding repetition patterns, navigation, description of symptom attributes. Another study pointed out five major drawbacks of AISCs based on users' perspectives: failing to factor into patient history in their algorithms, rigid input requirements, problematic probing questions, lacking support for all health conditions, and being short of sufficient functions for follow-up treatments [138].

Recent research has also reviewed the development process of healthcare applications, such as the applications' developers [72], regulation, consent and ownership [100], the role of health applications in healthcare [71]. Researchers found that there were a lack of professional medical involvement in the development process and design of mobile applications [19] and a lack of criteria to assess the expert or authority of the medical content [98]. These factors can influence the credibility of health applications. However, these studies all focus on the health applications market in general and do not specifically focus on AISC apps to explore their reliability and authority.

Little research has examined the factors influencing the authority of AISC apps from users' perspectives. To our best knowledge, only one study stated that symptom checkers could be viewed as a dubious authority as they may misdiagnose users [69]. Therefore, our study aims to explore the authority of AISC apps from users' perspectives.

### 2.2 Authority in medicine

When patients seek help from doctors, they usually grant authority and responsibility to doctors [127]. Patients make decisions not only due to the evaluation of relevant information, but also because of doctors' authority [68]. The authority of doctors can be regarded as a kind of expert authority (i.e., "authority grounded in a persistent differentiation of function and specialized ability between bearer and subject" [10]). Other studies regard doctors as a charismatic authority (i.e., "the qualities of an individual distinguished by a strong and magnetic personality that makes him/her attractive to a wide population and gives him the possibility to control and affect large") [80,96]. One study also recognizes doctors as an epistemic authority (i.e., because doctors "possess superior knowledge and judgment about diagnosis, prognosis, and the medical consequences of treatment", people believe in doctors' proposition because of their faith in doctors rather than the content) [6]. Medical authority is a key institution in medicine that structures the legitimacy of healthcare practice, the distribution and validation of medical knowledge, as well as certification and licensure [26,41]. Thus, the authority could engender trust, but differs trust that is focused on a type of belief in a one-to-one relationship [117]. In this paper, medical authority captures our interest in how AISC apps impact not only their users' trust, but also decision-making, action, experience, and broader socio-technical contexts.

Prior research has examined interior and exterior factors that would influence doctors' authority. The interior factors consist of doctors' communication skills in consultations [88,94], exclusive medical knowledge and expertise [45,47,64,80,83], reputation [83], and uniqueness of diagnostic results given by doctors [29]. Starr stated that doctor authority stemmed from the legitimacy of doctors ("which derives from their training, credentialing and expertise") and patients' belief on them [83]. Another study conducted in Lebanese has investigated people's trust in doctors [114]. This study found that good personal characteristics (e.g., "non-materialistic") and social factors (e.g., a good reputation) of doctors could stimulate more trust from the public. In addition, this study found that doctors graduating from reputable schools and working in respectable hospitals could also gain people's trust. The exterior factors include the appearance of doctors [55], governments' monitoring of medical quality [113], and societal characteristics [47]. Although acknowledging doctors as primary decision-making authority [59], patients sometimes challenge doctors' authority based on their own feelings [97], their rising educational levels [47], easier access to medical knowledge [80,99,132], and autonomy ("patients make their own decisions, and that they are enabled to do so") [126].

Information technology (IT) also influences how healthcare consumers understand doctors' authority [93]. Healthcare consumers grant cognitive authority (i.e., "the information sources, people or texts, that influence people's thoughts and are deemed credible" [51]) to relatable sources regardless of whether they are officially authoritative [44]. For example, patients searched for medical information not only from medical professionals but also from IT resources, such as online social networks or forums [39,133] and Internet searches [5,90,106,123]. Through online forums, patients tend to present themselves as authoritative, providing medical information for other people [7,51]. However, people do not always use information from IT to question doctors' authority; rather, they sometimes utilize online medical knowledge to affirm medical authority. For example, patients would still choose to see a doctor after searching for medical information online [105,120].Thus, IT might undermine the monopoly of doctors [5,14,18,21,49,53,125] through computerized retrieval and information storage [47,93,124], and change the doctor-patient relationship in medical consultations [46,48,82,83,110].

These studies have investigated the distribution of authority in medicine, but they mainly focus on offline medical visits and online information seeking. Little research has focused on the authority of consumer-facing AI-powered diagnostic systems. Also, it is still unclear what the difference is between these AI systems and human authority in medicine (i.e., doctors). Therefore, our paper explores the authority of consumer-facing AI health technologies and how they affect medical authority.

## 2.3 AI authority

AI systems can process information like human beings and aid complex tasks [107] through computer algorithms [56,122]. The use of AI could result in "the isolation of a particular information processing problem, the formulation of a computational theory for it, the construction of an algorithm that implements it, and a practical demonstration that the algorithm is successful" [79].

Several studies have examined the AI authority by comparing AI systems with human authority in a variety of settings, such as law, art, biosensing, driving, sports journalism, and education. A study found that artificial performers gained authority because they could supplement and even outperform human performers [62]. Suggestions provided by artificial performers seem to be professional and based on scientific evidence, making these artificial performers authoritative. Another example is that researchers illustrated that AI systems had the same authority as humans in monitoring cheating through an experimental study and questionnaire responses [52]. However, some people question the authority of AI systems especially regarding law [134] and art [54]. These studies demonstrated that AI systems could not replace human authority due to the lack of human value [54], morality, social intelligence, human judgment [42], and judicial transparency [8,15,121]. For instance, a robot lawyer could not replicate a human lawyer since only humans could deal with the core lawyering function [134]; AI art lacked the authority of creativity because people doubted its autonomy [54]; people showed their concern for the automated driving due to its unclear legal accountability [15].

Specifically, some AI researchers have drawn from the notion of algorithmic authority, which is "the legitimate power of algorithms to direct human action and to impact which information is considered true" [73], to discuss people's trust in AI algorithms [38,50]. Although AI algorithms can introduce risk and bias [16,135], the existing body of research has pointed out that people ascribe authority to AI-based systems because they tend to trust automated aids [34], believe in biodata [3], and anticipate that they can get better decisions from these systems [2]. Researchers also found that people trusted AI algorithms if they were efficient [65] and objective [37,65]. For instance, one study illustrated that users presumed that advertising algorithms are authoritative since they can receive appropriate advertisements [37].

With these concerns, previous studies have emphasized how to deal with the uncertain authority of AI systems and algorithms. First, researchers argued that AI algorithms or applications should involve the human intervention or human judgment [73,119,129]. Studies have pointed out that humans should always be in the decision-making loops [1] and be responsible for the final decision [116]. One study found that AI algorithms with the intervention of a human could outperform either of the two [67]. Concerning judicial transparency, having a human-in-the-loop can assign liability to the human authority and hold the human accountable [66]. Second, researchers found that people would increase the trust in AI algorithms only after the algorithms were well-developed and people understood their capabilities [50].

These studies have revealed the difference between AI authority and human authority as well as people's assessment of AI authority. Yet, we have limited knowledge of how AI authority is constructed in the wild. Also, although these studies have compared AI with human authority, they did not consider AI authority by entangling AI systems with all other stakeholders, such as human authority, users, and other IT systems. Nor do prior studies focus on the authority of AI in the domain of healthcare. Thus, our study aims to explore the medical authority of AI in healthcare.

## 3 METHODS

In this section, we will describe the AISC apps we studied in our work and the qualitative study conducted to investigate how our participants assessed medical authority.

### 3.1 AISC Apps under study

As the digital healthcare industry in China has developed rapidly, many AISC apps have emerged in China's healthcare market. The most well-known AISC apps among them are WeDoctor, Zuoshou Doctor, Chunyu Doctor, Mighty Doctor, and Intelligent Doctor. These apps are popular with relatively high ratings in mobile app stores (above 4.0/5.0). These apps are developed by either hospitals or for-profit companies backed by medical agencies. For example, Intelligent Doctor, a mini program in WeChat, is developed and operated by Guangdong Second Provincial General Hospital. Mighty Doctor, a mini program embedded in Alipay, was developed by the VHS group (Valurise Health Solutions International group, a company providing AI risk control services) and Mayo Clinic (an American nonprofit academic medical center) [139]. WeDoctor and Chunyu Doctor are developed by big technology companies. Symptom checkers usually use AI technologies to analyze big data and triage symptoms [86]. The AISC apps in our paper are embedded with AI technology as stated in their websites. For example, Zuoshou Doctor uses AI technologies such as deep learning, big data processing, and semantic comprehension with a conversational user interface [140].

Users usually self-diagnose themselves with an AISC app following three steps: creating a personal profile, inputting symptoms, and receiving an initial diagnosis. In the first step, most AISC apps require users to enter age (e.g., Intelligent Doctor, Mighty Doctor, Chunyu Doctor, and Zuoshou Doctor). Some apps also allow users to input other personal demographic information. For example, users can input information regarding their gender, weight, height, and medical history on Zuoshou Doctor.

In the second step, these apps utilize diverse forms to solicit input from users. Some apps (e.g., Zuoshou Doctor, WeDoctor, and Intelligent Doctor) interact with users and collect information through a chatbot (a computer program using Artificial Intelligence techniques, which can be used to interact and communicate with humans [70]). Other apps (e.g., Chunyu Doctor and Mighty Doctor) ask users to select their symptoms from a preset list of symptoms or a human figure (see Figure 1).

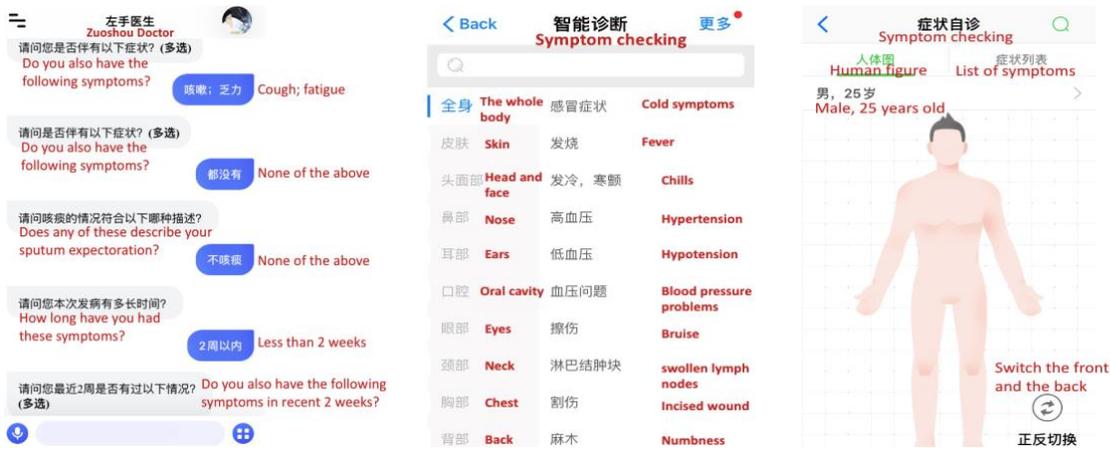

Figure 1: Different forms of inputting symptoms: the chatbot of Zuoshou Doctor (left), the symptom list of Mighty Doctor (middle), and the human figure of Chunyu Doctor (right).

In the last step, all these apps provide a diagnostic report. For example, Chunyu Doctor provides a report including tentative medical diagnoses, "similarity ratio" (i.e., the probability showing that how many people with the same symptoms as the participant were diagnosed with the potential disease), treatment options, and relevant recommendations (e.g., recommending online or offline doctors and hospitals) (see Figure 2).

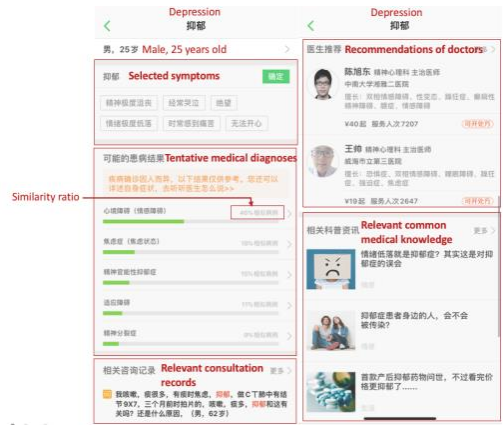

Figure 2: A diagnostic report of Chunyu Doctor.

## 3.2 Researcher reflexivity

We are university researchers from China or the US, working at the intersection of HCI and health informatics. We conduct research to help make the health IT systems support quality health service delivery at a lower cost, and support individuals to manage their own and their loved ones' health easily and safely. There is no conflict of interest between us and any health technology companies. The study was motivated primarily by the observation that many of our friends and family members in China and the US use AISC apps. Although some companies, such as Babylon (see [85,103]), claim that their AISC apps have high accuracy, it remains in doubt whether these tools are safe for users. Driven by this concern, we feel it is important to investigate how users perceive these symptom checkers.

## 3.3 Data collection and analysis

To understand how AISC apps affect medical authority from users' perception and experience, we conducted 30 semi-structured interviews with AISC app users from China. We obtained the university IRB approval before the study. We recruited participants through social media and snowball. Specifically, we distributed the flyer through WeChat Moments and Weibo (i.e., a popular Chinese microblogging website similar to Twitter). Our own social contacts also broadly distributed the flyer via their social media. Our participants had diverse backgrounds and did not know each other (see the Table 1 in Appendix). Some of our participants have some basic knowledge about AI as basic programming classes are usually required for college students in China, but most of them did not. An online screening survey was then sent to people who were interested in our study to establish eligibility. We collected information about their age, the AISC apps they had used, their goals of using these apps, and the last time they used the AISC app(s). The eligibility criteria specified that participants should be over the age of 18, had used AISC apps to consult for possible illnesses, and used AISC apps within one year. We stopped conducting interviews when the data reached "theoretical saturation [27]". We finally recruited 30 participants with diverse backgrounds. Our participants had diverse occupations, such as student, engineer, data analyst, landscape designer, and accountant. Their age ranged from 20 to 55. Most of them were in their 20s and 30s. They had used a variety of AISC apps for themselves or their family members or friends (e.g., S8 used Zuoshou Doctor for her mother). Then we conducted audio-recorded interviews via Wechat calls or Zoom videos chat in Mandarin with the consent of our participants. The interview duration ranged from 30 minutes to 1 hour. We started by asking their demographic information, when they started to use AISC apps, and how they started to use AISC apps. We then probed on their experiences of AISC app use, such as how they perceived the accuracy of diagnostic results from these apps, how they felt about the interactions with these apps, and whether they felt these apps were authoritative. We particularly asked which app they preferred if they had used multiple apps and how they perceived the conversation design of AISC apps if these apps use a chatbot to interact with users. We then transcribed these interviews in Mandarin.

We applied thematic analysis to analyze these transcripts in an inductive approach [17]. We first immersed ourselves in reading the data to familiarize the whole dataset. We then started to analyze data and generated over thirty initial codes using the data analysis software DeDoose. After that, we searched for the relationship between codes and formed candidate themes and sub-themes. Next, to reach internal homogeneity and external heterogeneity [75], we reviewed and refined these themes through rounds of discussions and coding. The final thematic map includes four parts: the endorsements by established authorities, data provenance and transparency, the charismatic authority through interaction design, and relational authority distributed in a socio-technical context of healthcare. When reporting findings, we use S1, S2, etc. to denote different participants. Representative quotes were translated into English for reporting.

## 4 FINDINGS

AISC apps impacted how our participants experienced and assessed medical authority. First, participants' authority perception of AISC apps was associated with established authorities. Second, authority perception was influenced by data provenance and transparency. Third, with ongoing interactions, participants' perception of AISC apps' authority was developed based on their interaction design patterns (i.e., "personalities"). Finally, we found the medical authority was distributed in an intertwined socio-technical context including all relevant stakeholders (e.g., the user self, human authority, search engines, online community, and other AISC apps).

## 4.1 The endorsements by established authorities

Users ascribe authority to AISC apps based on whether they are supported by established authorities. That is, the authority of AISC apps is reliant upon the reputation of the entities running behind them. For example, S10 told us:

"*First of all, you must have an understanding of this platform before you choose it, right? For example, I knew the background of Chunyu Doctor and Alipay from the Pencilnews or 36 Kr. Before using these apps, I have read news articles about them, and learned that they are reliable, because their developers are specialized to make these medical apps. Let alone the background of Alipay, which is a big company, you will naturally trust a big platform, a big company.*" (S10)

For this participant, she chose Chunyu Doctor and Mighty Doctor only after she had an understanding of their operator background. She approved the authority of these two AISC apps since they were developed by professional companies and thus had a reliable background endorsement.

Also, participants were likely to ascribe authority to AISC apps that were affiliated with hospitals or doctors, as S21 put:

"*On Baidu, everyone can give an answer to me. Thus, its information is not trustworthy. Chunyu Doctor and Zuoshou Doctor are official apps provided on platforms that also have online doctors. The co-presence of apps and online doctors definitely increases the credibility of these apps. I think Baidu is less trustworthy than searching on these professional platforms.*" (S21)

Baidu is a Chinese search engine similar to Google. Baidu does not have a medical knowledge base or database. People can search for their symptoms using Baidu and receive a diversity of answers from various websites. Thus, it is difficult to identify the credibility and accuracy of these answers and websites (e.g., S17 stated, "*You don't have the ability and time to distinguish the correct information from Baidu*"). Therefore, in the case of S21, authority was granted to AISC apps instead of Baidu due to AISC apps' affiliated online doctors.

Similarly, S7 tended to ascribe authority to AISC apps rather than an online community (i.e., Baidu Knows, which is a Chinese question-and-answer online community). She thought online peers exaggerated her conditions due to the lack of a certification. She commented:

"*I have used Baidu Knows before to ask questions. But the answer was too exaggerative. Some online users are not very responsible. They would say that you had very serious diseases, because they may not have a certification. So asking questions on Baidu Knows may not be so credible as AISC apps.*" (S7)

## 4.2 Data provenance and transparency

Authority perception is also influenced by AISC apps' data sources' credibility and transparency. Participants desired to know data provenance, or where the data originated from. Most of them regarded the data from established authorities as credible and authoritative. Thus, participants' assessment of AISC apps' authority was associated with the credibility of the data source. For instance, S10 mentioned:

"*If the app wants to increase the credibility, it should disclose some information appropriately, such as how many hospitals, medical records these apps used. The information presented on these platforms including doctors, hospitals, patients, and symptoms, etc., [can improve their credibility].*" (S10)

As S10 said, AISC apps could improve their credibility by showing that their data was provided by established authorities (e.g., hospitals). She desired more transparency such as whether these apps used medical records.

While the participant above expressed her desire to know the data source of AISC apps, other participants doubted the data source and data quantity of these apps. S12 told us:

"*I think their data [used to provide diagnoses] is insufficient. Because these apps were just developed. It's unclear where their data comes from, and if they have enough data. I feel hospitals wouldn't give their data to those companies, so I think their information and data is not enough. If they show me that their algorithmic models are obtained from big data mining, they may have to prove where the data comes from, whether the data is accurate and correct, right? This is very important. If the data is unreliable, it may affect their algorithmic models.*" (S12)

In this case, S12 cast doubt over the data source and data quantity of AISC apps. From her viewpoint, the accuracy of AISC apps was highly related to their data sources. Since she guessed that these AISC apps could not acquire data from the established authorities (i.e., hospitals), she questioned the accuracy of AISC apps and hesitated to ascribe authority to them.

Thus, from the viewpoints of the above participants, the authority of AISC apps was associated with the credibility of the data sources.

### 4.3 The charismatic authority of AISC apps developed through the interaction design patterns

Similar to doctors who maintain charismatic authority through their personalities and appearances such as the style of attire, AISC apps also acquire charismatic authority through their perceptible features (i.e., interaction design patterns). Our participants' assessment of AISC apps' authority was developed through their interaction with these apps, during which they evaluated AISC apps' input functions, probing questions, and anthropomorphic design.

*4.3.1 The charismatic authority of input design.*

User input is usually the first type of interaction in users' encounters with AISC apps. Its design helps participants form the first impression of AISC apps' authoritativeness. For example, participants would evaluate the authority of AISC apps based on the way they solicited symptom information. On the one hand, when participants felt that AISC apps were presenting them the right set of symptoms to select from, they would ascribe authority to AISC apps. As S21 put, "*I think the results given are correct, because, for example, it asked me some symptoms, which I think I also have, such as my dry throat and the like.*" (S21)

On the other hand, if participants could not input their symptoms in their expected way, they started to question the authority of AISC apps. For example, S12 was upset by the fact that the term "runny nose" was under the category of nasal congestion. Because she only had a runny nose without nasal congestion, she perceived the input function was insufficient to input her symptoms accurately, as she explained:

"*But I don't think the options it offers are comprehensive. For example, I know that I have caught a cold recently, and then I wanted to choose the symptom of runny nose, but I could not find it at the first glance. And after a long while, I found it in the category of nasal congestion. Then I felt that its options are not very comprehensive, because I didn't have the symptom of nasal congestion. It is not easy to use.*" (S12)

Additionally, the limited input also influenced participants' assessment of AISC apps' authority, as S16 put, "*It doesn't let me describe my symptoms. It just lets me choose from its options, to choose some symptoms that* *are plausible. So, the final result for me is only for reference.*" (S16) In this case, since S16 could only choose symptoms from a preset list but not describe them by text, she doubted the accuracy of her input, influencing her assessment of the authority of AISC apps.

*4.3.2 The charismatic authority of probing questions design.*

Users' second type of interaction with AISC apps is answering probing questions, which also impacts participants' assessment of AISC apps' authority. Participants tended to evaluate AISC apps' authority by examining if their probing questions were comprehensive and in-depth, as illustrated by S21's comment,

"*I think that Chunyu Doctor is incomprehensive, just like I selected 'snoring', and then it asked me if I had the following symptoms. I then selected 'snore', 'nasal congestion', and 'blocked nasal cavity' from its options. It didn't ask me how long these symptoms lasted, and when they started, it asked generic questions. I couldn't believe it, and I felt it was bad.*" (S21)

Drawing from his experience, lacking probing questions regarding frequency and duration of symptoms made him doubt the comprehensiveness of Chunyu Doctor, determining his assessment of its authority. Similar to S21's experience, S11's assessment of Zuoshou Doctor's authority was affected by the great number of irrelevant questions, as she complained:

"*I asked about the acne on my face, and then it [Zuoshou Doctor] asked me a lot of questions. It asked me what other symptoms I may have, I said no. Then it asked again, 'what other symptoms did you have,' and after I selected some symptoms, it continued to ask me, 'do you still have any other symptoms?' But in fact, I just have the acne on my face. It asked me so many questions!*" (S11)

In the above comments, S11 complained about the number of questions. She consulted for the acne on her face, but Zuoshou Doctor asked too many questions that were irrelevant to the acne. These inapplicable questions made her hesitate to grant authority to Zuoshou Doctor.

*4.3.3 The charismatic authority of anthropomorphic design.*

The anthropomorphic design might also influence participants' assessment of AISC apps' authority. Some participants desired human-like conversations. For instance, S9 noted that "*I think using human-like dialogue is better, as we can get better user experience and accept these apps more easily.*" (S9)

However, participants stressed that the conversational style should be serious in the medical context. They tended to ascribe authority to formal language rather than human-like language. For example, S10 said:

"*There is no need to be too personified. In that case, the app may look like cute. It would seem like the app has a child's tone, which would make me feel it untrustworthy naturally. Anthropomorphism may be good for some entertainment platforms, but not for medical-related platforms. The design of medical-related platforms should be more serious.*" (S10)

In this case, S10 thought since the AISC app was for medical-related diagnoses instead of entertainment, the language used in the app should be more formal and serious. The overly personified language would make the app seem less authoritative. Another participant stressed the accuracy of language use in his authority assessment. He told us:

"*For user experience, the human-like conversational style of Zuoshou Doctor is very good. But after all, the platform is used for medical consultation rather than for enjoyment; it is a platform for solving problems. The* *importance of user experience may be less than the importance of accuracy. I think the most important thing is accuracy, which is greater than user experience and its promotion.*" (S14)

In the above comments, S14 thought the anthropomorphic conversational design was acceptable and could improve the app's user experience. However, accuracy was a more important factor than user experience when considering authority because the AISC app was medical-related and aimed at problem-solving.

### 4.4 The relational medical authority distributed in a socio-technical network

Authority implies relational ties, and can be distributed over a group of agents in the field [32,58]. We found that participants ascribed medical authority not solely based on their interactions with one particular AISC app, but through a network of medical entities: they not only used one AISC app, but also utilized multiple channels, such as themselves, online doctors, offline doctors, online search engines, online communities, and various AISC apps for the purpose of cross-validation and comparison. In this network, the authority of AISC apps is thus relational, distributed across multiple entities.

*4.4.1 Relational medical authority arising from respect for diagnostic procedures.*

AISC apps' authority was also evaluated based on how these apps related themselves to the established diagnostic procedures. For example, AISC apps always showed respect for doctors or hospitals through their prompts, reflecting how they want to be accurate in their language and capacity, as S17 described, "*I think these AI diagnoses are still very conservative about the probability of their diagnoses, they show this kind of prompt on their platforms: 'If you really feel particularly unwell, you had better go to the hospital and call 120'.*"

Specifically, the authority of AISC apps is influenced by users' assessment of diagnostic procedures adopted by established authorities (i.e., established diagnostic procedures). Usually, people acquire an accurate diagnosis through complex diagnostic procedures in clinical practice. When using AISC apps, our participants tended to compare their procedures with the established ones. The relational authority is developed through this process.

To reach an accurate diagnostic result, an established diagnostic process needs to guarantee that sufficient information is collected, such as health history and lab test results. However, participants perceived that AISC apps could not collect sufficient information (e.g., body test data) in their probing processes comparing with the established diagnostic procedures, which affected their assessment of AISC apps' authority. S14 reported:

"*For some diseases that I don't know, or some severe diseases, I feel the app is very unreliable. Because when we go to the hospital, the doctor will also let us do a blood test first, do a CT scan, or a B-scan ultrasonography. Without these tests, even you consult a famous doctor, he or she cannot perform a good diagnosis.*" (S14)

In this comment, S14 did not grant authority to AISC apps because these apps could not do some lab tests, such as CT scan and B-scan ultrasonography that were necessary for clinical settings.

In addition, some participants found that AISC apps lacked expertise in developing a right order of probing questions, in comparison to the standard order of questions asked by a doctor during a clinical visit. The communication skills of medical professionals are important in the probing process, since various probing questions are associated with patients' feelings [91]. Thus, some participants were not satisfied with how AISC apps ask probing questions, realizing these AISC apps lacked the communication skills of human doctors. S16 reported:

"*I think the results from human doctors are more accurate. If we don't understand some medical jargon, or if we want to know more, human doctors can immediately understand what you want to say, and then guide you step by step to speak it out.*" (S16)

Here, this participant thought in established diagnostic procedures, human doctors could guide patients to describe their symptoms efficiently, while AISC apps could not do this. Also, from the standpoints of some participants, a human doctor had more comprehensive knowledge than AISC apps, as S12 said,

"*It's more reliable to go to hospitals and talk with the doctor face to face, because these apps or platforms only have explicit knowledge, but the doctors also have their tacit knowledge, and their own experience after diagnosing with a large number of patients.*" (S12)

This participant thought human doctors had not only explicit knowledge that could be used in AISC apps but also tacit knowledge which could only be acquired through previous experience. Since AISC apps did not have tacit knowledge as human doctors, S12 did not ascribe authority to AISC apps.

In addition, in a clinical visit, healthcare consumers know clearly that doctors and hospitals are held accountable for the diagnoses. However, AISC apps did not have a clear accountability mechanism. When our participants did not have a clear understanding of AISC apps' procedures and accountability mechanisms, their authority assessment was affected. S9 described:

"*When I went to the hospital to see a doctor, I have records here. If something goes wrong, the hospital will be responsible. But after I use an app, if there are some follow-up problems that are not easy to solve, how can I deal with them? The developer is far away from me. I don't know which hospital owns it, and I don't know if it will be accountable.*" (S9)

In this case, S9 complained that AISC apps did not disclose who should be responsible for follow-up problems after their medical consultations. In contrast, in a clinical visit, physicians or hospitals will be held accountable.

*4.4.2 Distributed medical authority through comparisons and cross-validations among entities.*

Participants constructed medical authority through clinical visits and the use of multiple diagnostic tools (e.g., AISC apps, Baidu, and online communities). They cross-validated the results from multiple diagnostic tools and referred to clinical diagnoses as well as their personal experience to make medical decisions. During these processes, the medical authority is distributed to different diagnostic tools, offline doctors, and the user self.

First, our participants cross-validated information from AISC apps and other online tools, including Baidu, other AISC apps, online communities, and Online Ask the Doctor platforms (i.e., platforms that allow people to consult a healthcare provider online). In other words, participants considered an AISC app authoritative when they confirmed its result with that from other tools. In turn, these tools were also considered authoritative. For example, S8 said, "*I think if I look for a diagnosis, I will not just use one software. I will definitely go to Baidu or something else. If I can get the same results, I might believe it.*" As she mentioned, she might ascribe authority to AISC apps only after she received the same results from AISC apps and Baidu. Likewise, S3 shared an experience of using Zhihu (a Chinese question-and-answer online community), Baidu, and AISC apps: "*Sometimes I also searched for answers on Zhihu again. After browsing a lot of information from Baidu, Zhihua and apps, I can confirm if I have any problems.*" In this case, S3 cross-validated the results from Zhihu, Baidu, and AISC apps when assessing the medical authority. Similarly, S17's authority assessment arose when different AISC apps presented to him the same diagnostic results based on similar inputs. he noted:

"*I also compared the results from these apps. Like you entered similar symptoms in these three apps. If they can give me similar diagnoses, then I believe that they are qualified, and their accuracy is acceptable.*" (S17)

In another case, S14 compared AISC apps with Online Ask the Doctor platforms, as he told us, "*When I got sick, I first asked online doctors, then I also used AI tools, so that I could have a comparison. It is similar to a comparative control experiment.*" Here, S14 used both AISC apps and Online Ask the Doctor platforms to cross-validate the diagnostic results.

Second, participants also referred to medical professionals to assess the authority of AISC apps. Some participants would grant authority to AISC apps if the results from AISC apps were similar to those from medical professionals. For example, S22 mentioned:

"*The first feeling it gave me is pretty good and reliable. It is consistent with my previous diagnosis by the doctor. I just wanted to see if it is reliable as the doctor said, if the results are the same, I will definitely use it in the future.*" (S22)

In her case, S22 tended to ascribe authority to the AISC app when she found the result from it was consistent with that from a previous medical visit.

If the results from AISC apps were different from the results diagnosed by medical professionals, participants would raise questions about AISC apps. For instance, S20 told us:

"*Zuoshou Doctor recommended me a kind of medicine. It said the medicine was used for joint pain and the like, which is quite consistent [with the patient's symptoms]. But when I went to the pharmacy, after I talked with the pharmacist, I thought I still had to see the doctor. The pharmacist said that there was just a similar one, and they said this medicine couldn't have a therapeutic effect. But the feedback from the Zuoshou Doctor did not make me feel that I must see a doctor. We plan to see the doctor later.*" (S20)

In this case, S20 used Zuoshou Doctor at first and planned to buy the recommended medicine because he was convinced by the AISC app. However, he changed his mind after talking with the pharmacist. The pharmacist recommended him a similar medicine and told him this medicine could not heal arthritis. As a result, S20 hesitated to ascribe authority to Zuoshou Doctor and decided to see a doctor instead.

Third, AISC apps, as a new factor in medical authority network, have influenced participants' decision making in clinical visits. Our study reflects that people came to AISC apps due to a variety of constraints, including physical, financial, and temporal. For instance, S23 used Chunyu Doctor because he lived in suburbia and it was difficult for him to go to a hospital, as he stated, "*I think Chunyu Doctor is good, because I can find medicine or something on it. I'm living in a remote place. It is inconvenient to go to the hospital. It may be more convenient to use the app for diagnosis.*" S15 showed his concerns for the high registration fees charged by the hospital, as he said, "*Many hospitals now charge between 10 and 50 RMB for registration fees. The fee to see a specialist is 300 RMB. It's hard for most family to pay it.*"

With the awareness of AISC apps' strengths, participants strategically used AISC apps as a tool in their clinical visits, distributing medical authority in both AISC apps and offline doctors. Before in-person visits, participants utilized AISC apps to judge whether they need to see a doctor, as 27 told us, "*For ailments, Zuoshou Doctor has a lot of detailed information, so there is no need to see a doctor. But I will go to see a doctor if my symptoms do not get better after a while or my situation is serious.*" In this case, S27 made decisions on whether to see a doctor based on the severity of her symptoms—she would go to the hospital only with serious conditions. Participants would also choose to see a doctor if all AISC apps suggested them to do it, as S13 told us, "*If all the results I got (from three apps) suggest that I need to go to the hospital, I will do it.*" However, not all participants heavily depended on AISC apps' recommendations to make decisions regarding whether they should see a doctor. For example, S29 said: "*Using Zuoshou Doctor has slight effect on whether to go to the hospital. Because if I feel my symptom is serious through my bodily experience, I will definitely go to the hospital no matter what results I get from Zuoshou Doctor.*" Here, S29 decided whether or not to see a doctor based on her bodily experience instead of recommendations from Zuoshou Doctor. In addition, some participants used AISC apps before the clinical visits in order to acquire symptom-related information and release mental pressure, as S21 said, "*I just took a look at what the disease might be using Zuoshou, so I would not be nervous, could have an overview of my symptoms, and could get prepared before talking to the doctor.*"

During clinical consultations, AISC apps were regarded as a helpful tool in the patient-provider communication. For example, S3 told us: "*I will tell the result from apps to the doctor, because my knowledge in medicine is very limited. I think these results are good references for doctors*." In this case, S3 thought diagnostic results from AISC apps were good references for doctors and thus could facilitate the communication.

After the clinical visits, some participants used AISC apps to seek for more possible diagnoses and information. For example, S8 told us:

"*After consulting a doctor, since the doctor did not give us an accurate diagnosis, we also used Zuoshou Doctor. It [Zuoshou Doctor] asked a few more questions, such as if my mother had numbness or nausea. I think there's a possibility of numbness because when I took my mother to the hospital, the doctor also asked if there were numbness-related symptoms. I think the diagnosis it offered is possible for my mother to have.*" (S8)

In this example, S8 used Zuoshou Doctor to explore possible diagnoses after seeing a doctor. This case implies the use of AISCs has changed the traditional practices of seeking diagnoses. Participants would use AISC apps as a supplement to clinical visits, which is in line with the previous research stating that people use IT to question or affirm doctors' authority. S20 reported a more specific example:

"*I twisted my leg when playing badminton, and then I went to the school hospital to check it. I felt that the doctor's diagnosis was not very accurate. I wanted to test it again. Because the plaster given to me by the doctor could only reduce swelling and analgesia. I wondered if there might be other medicines and if I had other diseases. I didn't think the doctor could give me more results, so I used the app to test it myself.*" (S20)

In this case, S20 came to Chunyu Doctor to seek for other possible medicines and medical diagnoses after seeing a school doctor, because S20 felt dissatisfied with the doctor's diagnosis and the plaster prescribed by the school doctor. Thus, the AISC apps, as a supplementary tool in clinical visits, have brought changes to medical authority.

Last, we found that participants would grant authority to AISC apps if the result was consistent with their bodily experience and perception. For example, S30 told to us:

"*I am 50 years old now, in menopause, and I can't sleep well. Zuoshou Doctor said it might be menopausal problems. I obviously have these menopausal symptoms now, so I think although Zuoshou Doctor is a machine, it's good at providing diagnoses*." (S30)

The above comments indicate that S30 evaluated the authenticity of Zuoshou Doctor based on her perception of her symptoms. Since Zuoshou Doctor diagnosed her illness as menopausal problems, which was consistent with her bodily experience, she tended to grant authority to Zuoshou Doctor.

However, if the results contradict participants' perception, they may lower their expectations for AISC apps' authority, as S15 said, "*I remember that Zuoshou Doctor said there was 90% chance of brain atrophy. It was impossible, because I think I am not that unhealthy. It's impossible for me to have such a serious illness.*" Here, since S15 thought the diagnostic result (i.e., brain atrophy) was inconsistent with his perception of his health status, he thought the diagnosis from Zuoshou Doctor seemed unreasonable and exagerative.

Participants also perceived the authority of AISCs by examining whether the features of AISCs are in line with participants' knowledge for the disease, as S11 said, "*In fact, before you ask questions, you already have some common knowledge. Then based on that, you can roughly judge what kind of disease you may have*." More specifically, S5 told us using his example: "*I have used multiple apps, such as Zuoshou Doctor and Chunyu Doctor. In fact, I know that people usually have hypoglycemia after playing basketball. Only Zuoshou Doctor accurately said I had hypoglycemia, which I thought was accurate*." Here, S5 thought the diagnostic result from Zuozhou Doctor (i.e., hypoglycemia) was accurate based on his knowledge for hypoglycemia—people may have hypoglycemia after playing basketball. Thus, S5 granted authority to Zuoshou Doctor based on his knowledge for the disease.

## 5 DISCUSSION

The present work aims at understanding how the advent of AISC apps influences people's assessment of medical authority in everyday healthcare practice. By invoking the notion of authority, our research was centered on the identification of socio-technical factors and processes that impacted how participants understood and constructed authority. Our research goal was not to provide a definitive, quantitative account of to what degree participants considered AISC apps authoritative. It is possible that regarding one particular feature on an AISC app, participants developed different authority perceptions. Such difference did not invalidate our findings.

We found that participants' medical authority assessment was influenced by the endorsements by established authorities, data provenance and transparency, and developed along with the interactions. Participants perceived and constructed authority based on the interaction design patterns of AISC apps. The medical authority was distributed in the socio-technical context of healthcare through the use of multiple self-diagnostic tools and comparisons with clinical visits. These findings together point to how AI helps distribute the medical authority within an ecosystem of multiple entities, and how medical authority is dynamically developed through human-AI interaction.

### 5.1 Distributing medical authority in a healthcare ecosystem

AI-enabled health technologies are a distinct force shaping the conception of medical authority in ordinary people's everyday health practices. On the one hand, the ways participants assessed the authority of AISCs were similar to how they evaluated doctors' personalities [88,94], their own feelings [97], or online searching using search engines [5,90,106,123]. They still acknowledged the medical authority of other entities in the healthcare ecosystem and prioritized the authoritativeness of established authorities like doctors and hospitals. Endorsements from established authorities like doctors, hospitals, or reputable healthcare companies could boost participants' confidence in those apps. Moreover, they relied upon similar patterns such as charismatic features to construct authority.

On the other hand, the authority conception of AISCs has unique characteristics. Previous studies have shown that healthcare consumers have utilized IT such as search engines for medical information seeking, which in turn shaped their authority perception [5,90]. While people have ascribed more authority to online information with reputable and credible sources or publishers [22,104], search engines are also criticized for lacking a medical database and the provision of unsubstantiated results, thus hurting their authoritativeness [112]. In comparison, AISC apps are usually supported by established authorities (i.e., hospitals and reputable companies). In addition, comparable to the epistemic authority of doctors who own trustworthy medical knowledge acquired from certified programs [64,80], our participants put more trust in AISC apps with reliable data sources, suggesting a credible knowledge base.

Different from previous AISC research that relies upon expert evaluations of AISCs [57,111], our work illustrates how users perceive and verify the medical authority of AI from a user's perspective. Existing studies often aim to develop a trust theory for healthcare chatbots [131] or to establish a standardized assessment framework to evaluate health AI technologies [108]. These studies focus on the criteria formulated to evaluate the healthcare applications. However, they do not illustrate how users perceive the applications in detail. Our study reveals that users perceive the medical authority of AI by comparing the recommendations from AISC apps with their bodily experience and medical knowledge, recommendations from other self-diagnosis tools (e.g., competing AISC apps and search engines), as well as diagnoses from offline doctors.

The medical authority of AI should be understood in terms of an ecosystem rather than a singular entity. In the *Anatomy of an AI System* [28], Crawford and Joler traced how the Amazon Echo relies upon its ecosystem, the global network of materials, labor, and data. Similarly, our participants' accounts of AISCs were ripe with how they were concerned about who developed the algorithms, who supplied the data, as well as who sponsored the apps. These are all constitutive elements that grant and sustain the medical authority of AISC apps—AISC apps are not produced in a vacuum.

There is also a broader ecosystem of medical authority to include entities that are not involved in their making. Recent studies have defined the healthcare "ecosystem" [95,128,137], including "physicians, patients, medical societies, hospital systems, software developers, the health information technology industry, and governmental regulatory agencies" [95]. However, these studies mainly focused on clinical practice rather than consumer-facing settings. Our study focused on the consumer-facing context of AISC apps, and showed that how participants distributed medical authority to AISC apps depended on its own features as much as other, unrelated entities (e.g., competing AISC apps, online communities, search engines), as well as unrelated established authorities. Thus, the medical authority of AI is a relational sociotechnical network including software developers, software operators, hospitals, medical professionals, healthcare consumers, online peers, search engines, and various online self-diagnosis tools.

The authority network influences AISC users' decision-making in two primary ways. First, the authority network renders the decision-making process distributed. That is, participants' medical decision-making does not rest with one single AISC app or offline doctors, but relies upon a socio-technical network of those aforementioned entities. Several participants mentioned how they moderated their medical authority perception through comparison with their personal experience and cross-validation with results from multiple AISC apps, search engines, online communities, as well as offline established authorities. A trend of previous research has found that users seek information from online peers [92], online social networks [39,133], and Internet searches [5]. Our work echoes with these studies and proposes a broader authority network regarding how patients seek for diagnostic information and make decisions in the practices of care management using AISCs.

Second, the authority network has potentials to enact AISC apps as a patient-centered system to supplement clinical visits and to break the "critical barriers between clinicians and patients [30]". In recent years, there is an increasing body of work in Health Informatics and HCI that emphasizes the importance of patient-centered care, revealing that health care systems should shift the focus to patients' needs [13,30], actively engage patients in the healthcare decision-making [118], and consider individual, health professions, as well as society to manage illness [43]. This calls for health technology to help healthcare consumers' self-management and explore their needs. As reflected in our findings, our participants would come to AISC apps when they met with temporal, physical, and financial constraints in offline medical visits. Therefore, AISC apps could be possibly regarded as a useful tool to support patient-centered care and supplement in-person visits.

Previous scholarship has shown how health technologies supported practices of care management (e.g., [9,23,33,61]). Similarly, the burgeoning AISC apps can be adopted in doctor-patient collaboration in everyday healthcare before, during, and after the clinical visits. Before in-person visits, our participants utilized AISC apps to acquire information of their symptoms, decide whether or not to see a doctor, and release mental pressure. This implies AISC apps could help patients prepare themselves with symptom-related information and release their mental pressure before seeing a doctor. During the offline visits, some participants mentioned AISC apps could be used as a kind of communication tool in the probing process. These two findings are in line with previous research that stated patient-centered health information technologies could help patients collect information and elicit their preferences to clinicians regarding health care decisions [30], reflecting that AISC apps have the potential to be regarded as a patient-centered health technology to help patients make medical decisions. After the clinical visits, participants utilized AISC apps to acquire other possible diagnoses. This reveals that AISC apps can be used as a supplement to clinical visits and influence doctor's authority, which is in line with a stream of previous studies reporting that people use information from IT to question or confirm doctors' authority [39,105,133]. Recent research has proposed the "last mile" of AI implementation, in which AI algorithms are tested and embedded in real-world practice [25]. As resesarchers emphasized the importance of crossing over technical challenges (e.g., data governance [20]) to bridge the "last mile" gap between the reality and AI algorithms, our work focuses on end users' viewpoints regarding how AI technology has been implemented in their daily life, illustrating how AI technology plays around in healthcare delivery. We reinforce the call for further research regarding how to bridge the "last mile" gap of AI technology to fit in healthcare users' needs.

Ascribing authority to AI healthcare technologies might come with risks for health consumers. Different from doctors, who own authoritative medical knowledge [64], the reliability of AISC apps is unclear [58,61]. While medical professionals possess a systematic body of medical knowledge to assess the efficiency and authority of clinical diagnostic tools [2,6,9], healthcare consumers usually do not have such knowledge. This is partially why they relied upon a broader ecosystem to assess AISC apps. Previous work has reflected upon how the authority of IT might mislead healthcare consumers [40,46], and how the inaccurate evaluation of AI may lead to severe consequences such as mortality [76,81]. Thus, there is an urgent need to understand the unintended consequences of AI health technologies as AI gradually transforms how people experience and perceive medical authority in their everyday healthcare practices.

### 5.2 Developing medical authority through human-AI interaction

Previous studies have evaluated AI systems' capabilities and outputs [62,134], such as accuracy and efficiency [37,65], autonomy [54], and ethics and honesty [42]. By contrast, our study explored how healthcare consumers' AI authority perception unfolds in human-AI interactions. In other words, it is not only the results, but also human-AI interaction processes, that matter to authority perception.

HCI researchers have already drawn from classic HCI knowledge such as usability guidelines and heuristics to design guidelines for human-AI interaction (e.g., [4,130]). Our research on medical authority contributes to this line of thought by articulating important dimensions to consider when designing human-AI interaction in the medical domain. First, charismatic design matters to healthcare consumers' authority assessment and trust, but not in a similar way they experience that of other AI applications such as chatbots. Previous studies emphasized how anthropomorphic design, such as anthropomorphic behaviors [78], active listening skills [136], and casual conversational style [60], could make chatbots charismatic. However, the anthropomorphic design could erode the authority of AISC apps. Our participants like S10 thought since AISC apps were health-related, too personified language would make them look less authoritative. Instead, participants preferred charismatic features that could indicate authority, such as using formal and serious language.

In addition, healthcare consumers' authority perception is not necessarily static. It might change along through healthcare consumers' medical journey in both offline and online worlds. For instance, after consulting with a pharmacy doctor whose statements were inconsistent with the results acquired from Zuoshou Doctor, S20's authority assessment of Zuoshou Doctor had been changed. Therefore, human-AI interaction is not only about the exact moment and the exact location where the interactions happen. Instead, the broader socio-technical context of human-AI interaction matters for us to understand how healthcare consumers experience AI health technologies.

### 5.3 Implications for design

Our work stressed the need for improving the design of consumer-facing AI-enabled healthcare systems. Designers and developers of AI-enabled healthcare systems should consider enhancing transparency by disclosing information from other entities in the AI ecosystem. The first type of information is the endorsements by established authorities. Consumer-facing AI-enabled healthcare systems can display information regarding their developers, sponsors, and data sources. In addition, Consumer-facing AI-enabled healthcare systems may present diagnostic information from other sources for reference to cross-validate medical authority. For example, relevant information from Baidu can be presented on the apps. Also, online or offline doctors' diagnoses of similar cases may be shown in the final report of AI healthcare systems.

AI-enabled healthcare systems may improve their interaction design, focusing on design patterns and procedural design. For example, in the case of AISC apps, they can improve their input flexibility, advance the presentation of probing questions, and intensify their knowledge base by mimicking established authoritative diagnostic procedures to gain authority. In the future, AI-enabled healthcare systems may provide guidance and warnings to let users notice the potential risks since they do not have sufficient knowledge to identify proper diagnostic information. AI-enabled healthcare systems can also introduce doctor-like customer services. When users doubt the diagnoses or have difficulties in describing their symptoms, they can turn to doctors for help.

## 6 CONCLUSION

Our study investigated how AISC apps influence users' assessment of medical authority. By conducting 30 semi-structured interviews, we described how medical authority was constructed along with the interactions between healthcare consumers and AI-enabled healthcare technologies. With the growing capacity and prevalence of AI in the healthcare domain, what we understand as medical authority is and will be experiencing tremendous changes. Even if the medical authority of AI systems today is still considered inferior to established authorities, we shall not take the impact of AI lightly. In healthcare, authority perception can have significant impacts on healthcare consumers' information-seeking, decision-making, behaviors, and ultimately health outcomes. As consumer-facing AI health technologies enter the playing field and influencing the landscape of medical authority, limited understandings of this influence have been accumulated. Future work should ponder how the medical authority of AI intersects with growing concerns over AI accountability, transparency, fairness, and ethics, as well as necessary regulatory frameworks to govern the authority landscape.

## ACKNOWLEDGMENTS

This work is supported by Penn State College of IST's seed grant, No.150000004308 INTR. We thank all the participants for sharing their experiences, and the reviewers at CHI'2021 for their constructive and insightful comments.

## REFERENCES

[1] Hussein A Abbass. 2019. Social integration of artificial intelligence: functions, automation allocation logic and human-autonomy trust. *Cognit. Comput.* 11, 2 (2019), 159–171.

[2] Ajay Agrawal, Joshua S Gans, and Avi Goldfarb. 2019. Exploring the impact of artificial Intelligence : Prediction versus judgment. *Inf. Econ. Policy* 47, (2019), 1–6. DOI:https://doi.org/10.1016/j.infoecopol.2019.05.001

[3] Miquel Alfaras, Vasiliki Tsaknaki, Pedro Sanches, Charles Windlin, Muhammad Umair, Corina Sas, and Kristina Höök. 2020. From Biodata to Somadata. In *Proceedings of the 2020 CHI Conference on Human Factors in Computing Systems*, 1–14.

[4] Saleema Amershi, Dan Weld, Mihaela Vorvoreanu, Adam Fourney, Besmira Nushi, Penny Collisson, Jina Suh, Shamsi Iqbal, Paul N Bennett, and Kori Inkpen. 2019. Guidelines for human-AI interaction. In *Proceedings of the 2019 chi conference on human factors in computing systems*, 1–13.

[5] James G Anderson, Michelle R Rainey, and Gunther Eysenbach. 2003. The impact of CyberHealthcare on the physician–patient relationship. *J. Med. Syst.* 27, 1 (2003), 67–84.

[6] Arthur Isak Applbaum. 2017. The Idea of Legitimate Authority in the Practice of Medicine. *AMA J. Ethics* 19, 2 (2017), 207–213.

[7] Natalie Armstrong, Nelya Koteyko, and John Powell. 2012. 'Oh dear, should I really be saying that on here?': Issues of identity and authority in an online diabetes community. *Health (Irvine. Calif).* 16, 4 (2012), 347–365. DOI:https://doi.org/10.1177/1363459311425514

[8] Susan C Athey, Kevin A Bryan, and Joshua S Gans. 2020. The allocation of decision authority to human and artificial intelligence. In *AEA Papers and Proceedings*, 80–84.

[9] Andrea M Barbarin, Laura R Saslow, Mark S Ackerman, and Tiffany C Veinot. 2018. Toward health information technology that supports overweight/obese women in addressing emotion-and stress-related eating. In *Proceedings of the 2018 CHI Conference on Human Factors in Computing Systems*, 1–14.

[10] Kenneth Benne. 1970. Authority in Education. *Harv. Educ. Rev.* 40, 3 (1970), 385–410.

[11] Chantal Benoit-Barné and François Cooren. 2009. The accomplishment of authority through presentification: How authority is distributed among and negotiated by organizational members. *Manag. Commun. Q.* 23, 1 (2009), 5–31.

[12] AC Berry, BD Cash, B Wang, MS Mulekar, AB Van Haneghan, K Yuquimpo, A Swaney, MC Marshall, and WK Green. 2019. Online symptom checker diagnostic and triage accuracy for HIV and hepatitis C. *Epidemiol. Infect.* 147, (2019). DOI:https://doi.org/10.1038/ajg.2017.305.Cite

[13] Clement Bezold, J Peck, W Rowley, and Marsha Rhea. 2004. Patient-centered care 2015: Scenarios, vision, goals & next steps. *Camden, ME Pick. Inst.* (2004).

[14] Wu Bing and Zhang Chenyan. 2014. Advances in Online Authority Research. *Appl. Mech. Mater.* 596, (2014), 994–997. DOI:https://doi.org/10.4028/www.scientific.net/AMM.596.994

[15] Francesco Biondi, Ignacio Alvarez, and Kyeong-Ah Jeong. 2019. Human-vehicle cooperation in automated driving: A multidisciplinary review and appraisal. *Int. J. Human–Computer Interact.* 35, 11 (2019), 932–946. DOI:https://doi.org/10.1080/10447318.2018.1561792

[16] Matthew Boyd and Nick Wilson. 2017. Rapid developments in artificial intelligence: how might the New Zealand government respond? *Policy Q.* 13, 4 (2017).

[17] Virginia Braun and Victoria Clarke. 2006. Using thematic analysis in psychology. *Qual. Res. Psychol.* 3, 2 (2006), 77–101.

[18] Alex Broom. 2005. Virtually he@ lthy: the impact of internet use on disease experience and the doctor-patient relationship. *Qual. Health Res.* 15, 3 (2005), 325–345. DOI:https://doi.org/10.1177/1049732304272916

[19] Arthur Willem Gerard Buijink, Benjamin Jelle Visser, and Louise Marshall. 2013. Medical apps for smartphones: lack of evidence undermines quality and safety. *BMJ Evidence-Based Med.* 18, 3 (2013), 90–92.

[20] Federico Cabitza, Andrea Campagner, and Clara Balsano. 2020. Bridging the "last mile" gap between AI implementation and operation:"data awareness" that matters. *Ann. Transl. Med.* 8, 7 (2020).

[21] Rajesh Chandwani and Vaibhavi Kulkarni. 2016. Who's the Doctor? Physicians' Perception of Internet Informed Patients in India. In *Proceedings of the 2016 CHI Conference on Human Factors in Computing Systems*, 3091–3102.

[22] Chuanfu Chen, Yuan Yu, Qiong Tang, Kuei Chiu, Yan Rao, Xuan Huang, and Kai Sun. 2012. Assessing the authority of free online scholarly information. *Scientometrics* 90, 2 (2012), 543–560. DOI:https://doi.org/10.1007/s11192-011-0524-5

[23] Chia-Fang Chung, Kristin Dew, Allison Cole, Jasmine Zia, James Fogarty, Julie A Kientz, and Sean A Munson. 2016. Boundary negotiating artifacts in personal informatics: Patient-provider collaboration with patient-generated data. In *Proceedings of the 19th ACM Conference on Computer-Supported Cooperative Work & Social Computing*, 770–786.

[24] Enrico Coiera. The fate of medicine in the time of AI Transforming the landscape of liver disease in the UK. DOI:https://doi.org/10.1016/S0140-6736(18)31925-1

[25] Enrico Coiera. 2019. The last mile: where artificial intelligence meets reality. *J. Med. Internet Res.* 21, 11 (2019), e16323.

[26] Peter Conrad and Joseph W Schneider. 1990. Professionalization, monopoly, and the structure of medical practice. *Sociol. Heal. illness, Crit. Perspect.* (1990).

[27] Juliet Corbin and Anselm Strauss. 2014. *Basics of qualitative research techniques and procedures for developing grounded theory*. Sage publications.

[28] Kate Crawford and Vladan Joler. 2018. Anatomy of an AI System. *Retrieved Sept.* 18, (2018), 2018.

[29] Leonie van Dam, Ernst J Kuipers, Ewout W Steyerberg, Monique E van Leerdam, and Inez D de Beaufort. 2013. The price of autonomy: should we offer individuals a choice of colorectal cancer screening strategies? *Lancet Oncol.* 14, 1 (2013), e38–e46. DOI:https://doi.org/10.1016/S1470-2045(12)70455-2

[30] Stuti Dang. 2018. Shared Decision Making-The Pinnacle of Patient-Centered Care. *J. Indian Acad. Geriatr.* 14, (2018).

[31] Benjamin Marshall Davies, Colin Fraser Munro, and Mark RN Kotter. 2019. A Novel Insight Into the Challenges of Diagnosing Degenerative Cervical Myelopathy Using Web-Based Symptom Checkers. *J. Med. Internet Res.* 21, 1 (2019), e10868. DOI:https://doi.org/10.2196/10868

[32] Lieke Van Deinsen and Beatrijs Vanacker. 2019. Found through translation: Female translators and the construction of 'relational authority'in the eighteenth-century Dutch Republic. *Early Mod. Low Ctries.* 3, 1 (2019), 60–80. DOI:https://doi.org/10.18352/emlc.90

[33] Xianghua Ding, Xinning Gui, Xiaojuan Ma, Zhaofei Ding, and Yunan Chen. 2020. Getting the Healthcare We Want: The Use of Online " Ask the Doctor " Platforms in Practice. In *Proceedings of the 2020 CHI Conference on Human Factors in Computing Systems*, 1–13.

[34] Yogesh K Dwivedi, Laurie Hughes, Elvira Ismagilova, Gert Aarts, Crispin Coombs, Tom Crick, Yanqing Duan, Rohita Dwivedi, John Edwards, and Aled Eirug. 2019. Artificial Intelligence (AI): Multidisciplinary perspectives on emerging challenges, opportunities, and agenda for research, practice and policy. *Int. J. Inf. Manage.* (2019), 101994.

[35] Upol Ehsan, Pradyumna Tambwekar, Larry Chan, Brent Harrison, and Mark O Riedl. 2019. Automated rationale generation: a technique for explainable AI and its effects on human perceptions. In *Proceedings of the 24th International Conference on Intelligent User Interfaces*, 263–274.

[36] Malin Eiband, Sarah Theres Völkel, Daniel Buschek, Sophia Cook, and Heinrich Hussmann. 2019. When people and algorithms meet: User-reported problems in intelligent everyday applications. In *Proceedings of the 24th International Conference on Intelligent User Interfaces*, 96–106.

[37] Motahhare Eslami, Sneha R Krishna Kumaran, Christian Sandvig, and Karrie Karahalios. 2018. Communicating Algorithmic Process in Online Behavioral Advertising. In *Proceedings of the 2018 CHI conference on human factors in computing systems*, 1–13.

[38] Rino Falcone and Cristiano Castelfranchi. 2009. Socio-Cognitive Model of Trust. In *Human Computer Interaction: Concepts, Methodologies, Tools, and Applications*. IGI Global, 2316–2323.

[39] Nicholas J Fox, Katie J Ward, and Alan J O'Rourke. 2005. The 'expert patient': empowerment or medical dominance? The case of weight loss, pharmaceutical drugs and the Internet. *Soc. Sci. Med.* 60, 6 (2005), 1299–1309. DOI:https://doi.org/10.1016/j.socscimed.2004.07.005

[40] Hamish Fraser, Enrico Coiera, and David Wong. 2018. Safety of patient-facing digital symptom checkers. *Lancet* 392, 10161 (2018), 2263–2264. DOI:https://doi.org/10.1016/S0140-6736(18)32819-8

[41] Eliot Freidson. 1974. *Professional dominance: The social structure of medical care*. Transaction Publishers.

[42] Yair Galily. 2018. Artificial intelligence and sports journalism: Is it a sweeping change? *Technol. Soc.* 54, (2018), 47–51. DOI:https://doi.org/10.1016/j.techsoc.2018.03.001

[43] Trisha Greenhalgh. 2009. Patient and public involvement in chronic illness: beyond the expert patient. *Bmj* 338, (2009).

[44] Devon Greyson. 2017. Health information practices of young parents. *J. Doc.* (2017), 778–802. DOI:https://doi.org/10.1108/JD-07-2016-0089

[45] Li Guo, Bo Jin, Cuili Yao, Haoyu Yang, Degen Huang, and Fei Wang. 2016. Which doctor to trust: a recommender system for identifying the right doctors. *J. Med. Internet Res.* 18, 7 (2016), e186. DOI:https://doi.org/10.2196/jmir.6015

[46] Michael Hardey. 1999. Doctor in the house: the Internet as a source of lay health knowledge and the challenge to expertise. *Sociol. Health Illn.* 21, 6 (1999), 820–835.

[47] Marie R Haug. 1976. The Erosion of Professional Authority : A Cross-Cultural Inquiry in the Case of the Physician. *Milbank Mem. Fund Q. Health Soc.* (1976), 83–106.

[48] A Herrmann-Werner, H Weber, T Loda, KE Keifenheim, R Erschens, SC Mölbert, C Nikendei, S Zipfel, and K Masters. 2019. "But Dr Google said…"--Training medical students how to communicate with E-patients. *Med. Teach.* 41, 12 (2019), 1434–1440. DOI:https://doi.org/10.1080/0142159X.2018.1555639

[49] Aaron Hess. 2012. Health, Risk and Authority in a Dysfunctional World: Online Ecstasy User Discourse. *Ohio Commun. J.* 50, (2012), 1–30.

[50] Pamela J Hinds, Teresa L Roberts, and Hank Jones. 2004. Whose job is it anyway? A study of human-robot interaction in a collaborative task. *Human–Computer Interact.* 19, 1–2 (2004), 151–181.

[51] Noora Hirvonen, Alisa Tirroniemi, and Terttu Kortelainen. 2019. The cognitive authority of user-generated health information in an online forum for girls and young women. *J. Doc.* (2019).

[52] Guy Hoffman, Jodi Forlizzi, Shahar Ayal, Aaron Steinfeld, John Antanitis, Guy Hochman, Eric Hochendoner, and Justin Finkenaur. 2015. Robot presence and human honesty: Experimental evidence. In *10th ACM/IEEE International Conference on Human-Robot*

*Interaction (HRI)*, IEEE, 181–188.

[53] Sierra Holland. 2019. Constructing queer mother-knowledge and negotiating medical authority in online lesbian pregnancy journals. *Sociol. Health Illn.* 41, 1 (2019), 52–66. DOI:https://doi.org/10.1111/1467-9566.12782

[54] Joo-wha Hong. 2018. Bias in Perception of Art Produced by Artificial Intelligence. In *International Conference on Human-Computer Interaction*, 290–303. DOI:https://doi.org/10.1007/978-3-319-91244-8

[55] David J Hutson. 2013. "Your body is your business card": Bodily capital and health authority in the fitness industry. *Soc. Sci. Med.* 90, (2013), 63–71. DOI:https://doi.org/10.1016/j.socscimed.2013.05.003

[56] Marcus Hutter. 2012. One Decade of Universal Artificial Intelligence. In *Theoretical foundations of artificial general intelligence*. Springer, 67--88.

[57] Sabin Kafle, Penny Pan, Ali Torkamani, Stevi Halley, John Powers, and Hakan Kardes. 2018. Personalized symptom checker using medical claims. In *HealthRecSys@ RecSys*, 13–17.

[58] William A Kahn and Kathy E Kram. 1994. Authority at work: Internal models and their organizational consequences. *Acad. Manag. Rev.* 19, 1 (1994), 17–50.

[59] Rabia Kahveci, Duygu Ayhan, Pınar Döner, Fatma Gökşin Cihan, and Esra Meltem Koç. 2014. Shared decision-making in pediatric intensive care units: a qualitative study with physicians, nurses and parents. *Indian J. Pediatr.* 81, 12 (2014), 1287–1292. DOI:https://doi.org/10.1007/s12098-014-1431-6

[60] Soomin Kim, Joonhwan Lee, and Gahgene Gweon. 2019. Comparing data from chatbot and web surveys: Effects of platform and conversational style on survey response quality. In *Proceedings of the 2019 CHI Conference on Human Factors in Computing Systems*, 1–12.

[61] Predrag Klasnja, Andrea Hartzler, Christopher Powell, and Wanda Pratt. 2011. Supporting cancer patients' unanchored health information management with mobile technology. In *AMIA Annual Symposium Proceedings*, American Medical Informatics Association, 732.

[62] Lenneke Kuijer and Elisa Giaccardi. 2018. Co-performance: Conceptualizing the role of artificial agency in the design of everyday life. In *Proceedings of the 2018 CHI Conference on Human Factors in Computing Systems*, 1–13.

[63] Liliana Laranjo, Adam G Dunn, Huong Ly Tong, Ahmet Baki Kocaballi, Jessica Chen, Rabia Bashir, Didi Surian, Blanca Gallego, Farah Magrabi, and Annie YS Lau. 2018. Review conversational agents in healthcare: a systematic review. *J. Am. Med. Informatics Assoc.* 25, 9 (2018), 1248–1258. DOI:https://doi.org/10.1093/jamia/ocy072

[64] Stephen R Latham. 2002. Medical Professionalism. *Mt Sinai J Med* 69, (2002), 363–9.

[65] Min Kyung Lee. 2018. Understanding perception of algorithmic decisions: Fairness, trust, and emotion in response to algorithmic management. *Big Data Soc.* 5, 1 (2018), 2053951718756684. DOI:https://doi.org/10.1177/2053951718756684

[66] Charlene Liew. 2018. The future of radiology augmented with artificial intelligence: a strategy for success. *Eur. J. Radiol.* 102, (2018), 152–156. DOI:https://doi.org/10.1016/j.ejrad.2018.03.019

[67] Lanny Lin and Michael A Goodrich. 2015. Sliding autonomy for UAV path-planning: adding new dimensions to autonomy management. In *Proceedings of the 2015 International Conference on Autonomous Agents and Multiagent Systems*, 1615–1624.

[68] Zhicheng Liu and Jeffrey Heer. 2014. The effects of interactive latency on exploratory visual analysis. *IEEE Trans. Vis. Comput. Graph.* 20, 12 (2014), 2122–2131. DOI:https://doi.org/10.1109/TVCG.2014.2346452

[69] Amber Loos. 2013. Cyberchondria: too much information for the health anxious patient? *J. Consum. Health Internet* 17, 4 (2013), 439–445. DOI:https://doi.org/10.1080/15398285.2013.833452

[70] Joao Luis, Zeni Montenegro, Cristiano André, and Rosa Righi. 2019. Survey of conversational agents in health. *Expert Syst. Appl.* 129, (2019), 56–67. DOI:https://doi.org/10.1016/j.eswa.2019.03.054

[71] Deborah Lupton. 2014. Apps as Artefacts: Towards a Critical Perspective on Mobile Health and Medical Apps. *Societies* 4, 4 (2014), 606–622. DOI:https://doi.org/10.3390/soc4040606

[72] Deborah Lupton and Annemarie Jutel. 2015. 'It's like having a physician in your pocket!'A critical analysis of self-diagnosis smartphone apps. *Soc. Sci. Med.* 133, (2015), 128–135. DOI:https://doi.org/10.1016/j.socscimed.2015.04.004

[73] Caitlin Lustig and Bonnie Nardi. 2015. Algorithmic authority: The case of Bitcoin. In *2015 48th Hawaii International Conference on System Sciences*, 743--752.

[74] Caitlin Lustig, Katie Pine, Bonnie Nardi, Lilly Irani, Min Kyung Lee, Dawn Nafus, and Christian Sandvig. 2016. Algorithmic authority: the ethics, politics, and economics of algorithms that interpret, decide, and manage. In *Proceedings of the 2016 CHI Conference Extended Abstracts on Human Factors in Computing Systems*, 1057–1062.

[75] Courtney Rees Lyles, Lynne T Harris, Tung Le, Jan Flowers, James Tufano, Diane Britt, James Hoath, Irl B Hirsch, Harold I Goldberg, and James D Ralston. 2011. Qualitative evaluation of a mobile phone and web-based collaborative care intervention for patients with type 2 diabetes. *Diabetes Technol. Ther.* 13, 5 (2011), 563–569. DOI:https://doi.org/10.1089/dia.2010.0200

[76] Carl Macrae. 2019. Governing the safety of artificial intelligence in healthcare. *BMJ Qual. Saf.* 28, 6 (2019), 495–498. DOI:https://doi.org/10.1136/bmjqs-2019-009484

[77] Luis Marco-Ruiz, Erlend Bønes, Estela de la Asunción, Elia Gabarron, Juan Carlos Aviles-Solis, Eunji Lee, Vicente Traver, Keiichi Sato, and Johan G Bellika. 2017. Combining multivariate statistics and the think-aloud protocol to assess Human-Computer Interaction barriers in symptom checkers. *J. Biomed. Inform.* 74, (2017), 104–122. DOI:https://doi.org/10.1016/j.jbi.2017.09.002

[78] Natascha Mariacher, Stephan Schlögl, and Alexander Monz. 2020. Investigating Perceptions of Social Intelligence in Simulated Human-Chatbot Interactions. In *Progresses in Artificial Intelligence and Neural Systems*. Springer, 513–529.

[79] David Marr. 1977. Artificial intelligence—a personal view. *Artif. Intell.* 9, 1 (1977), 37–48.

[80] Tom Marshall. 1997. Scientific knowledge in medicine: a new clinical epistemology? *J. Eval. Clin. Pract.* 3, 2 (1997), 133–138. DOI:https://doi.org/10.1046/j.1365-2753.1997.00075.x

[81] Margaret Mccartney. 2018. Margaret McCartney: AI in medicine must be rigorously tested. *Bmj* 361, (2018), k1752. DOI:https://doi.org/10.1136/bmj.k1752

[82] John B Mckinlay and Lisa D Marceau. 2002. The end of the golden age of doctoring. *Int. J. Heal. Serv.* 32, 2 (2002), 379–416.

[83] Alka V Menon. 2017. Do online reviews diminish physician authority? The case of cosmetic surgery in the US. *Soc. Sci. Med.* 181, (2017), 1–8. DOI:https://doi.org/10.1016/j.socscimed.2017.03.046

[84] Ashley ND Meyer, Traber D Giardina, Christiane Spitzmueller, Umber Shahid, Taylor MT Scott, and Hardeep Singh. 2020. Patient Perspectives on the Usefulness of an Artificial Intelligence–Assisted Symptom Checker: Cross-Sectional Survey Study. *J. Med. Internet Res.* 22, 1 (2020), e14679. DOI:https://doi.org/10.2196/14679

[85] Katherine Middleton, Mobasher Butt, Nils Hammerla, Steven Hamblin, Karan Mehta, and Ali Parsa. 2016. Sorting out symptoms: design and evaluation of the'babylon check'automated triage system. *arXiv Prepr. arXiv1606.02041* (2016).

[86] S Miller, S Gilbert, V Virani, and P Wicks. 2020. Patients' utilisation and perception of an AI based symptom assessment and advice technology in a British primary care waiting room: Exploratory pilot study. *JMIR Hum. Factors* (2020).

[87] Stephen Miller, Stephen Gilbert, Vishaal Virani, and Paul Wicks. 2020. Patients' Utilization and Perception of an Artificial Intelligence--Based Symptom Assessment and Advice Technology in a British Primary Care Waiting Room: Exploratory Pilot Study. *JMIR Hum. Factors* 7, 3 (2020), e19713. DOI:https://doi.org/10.2196/19713

[88] Philip J Moore, Amy E Sickel, Jennifer Malat, David Williams, James Jackson, and Nancy E Adler. 2004. Psychosocial factors in medical and psychological treatment avoidance: The role of the doctor--patient relationship. *J. Health Psychol.* 9, 3 (2004), 421–433. DOI:https://doi.org/10.1177/1359105304042351

[89] Richard Munassi. 2020. 8 things to know about online symptom checker applications. *Becker's Healthcare*. Retrieved from https://www.beckershospitalreview.com/healthcare-information-technology/8-things-to-know-about-online-symptom-checker-applications.html

[90] Sarah Nettleton, Roger Burrows, and Lisa O Malley. 2005. The mundane realities of the everyday lay use of the internet for health,

and their consequences for media convergence. *Sociol. Health Illn.* 27, 7 (2005), 972–992. DOI:https://doi.org/10.1111/j.1467-9566.2005.00466.x

[91] Mary K O'Brien, Keith Petrie, and John Raeburn. 1992. Adherence to medication regimens: updating a complex medical issue. *Med. Care Rev.* 49, 4 (1992), 435--454.

[92] Aisling Ann O'Kane, Sun Young Park, Helena Mentis, Ann Blandford, and Yunan Chen. 2016. Turning to peers: integrating understanding of the self, the condition, and others' experiences in making sense of complex chronic conditions. *Comput. Support. Coop. Work* 25, 6 (2016), 477–501.

[93] Christopher Pearce, Michael Arnold, Christine Phillips, Stephen Trumble, and Kathryn Dwan. 2011. The patient and the computer in the primary care consultation. *J. Am. Med. Informatics Assoc.* 18, 2 (2011), 138–142. DOI:https://doi.org/10.1136/jamia.2010.006486

[94] Anssi Perakyla. 1998. Authority and accountability: The delivery of diagnosis in primary health care. *Soc. Psychol. Q.* (1998), 301–320.

[95] Filippo Pesapane, Marina Codari, and Francesco Sardanelli. 2018. Artificial intelligence in medical imaging: threat or opportunity? Radiologists again at the forefront of innovation in medicine. *Eur. Radiol. Exp.* 2, 1 (2018), 35.

[96] Anja Petaros, Č Mislav, Andrea Šuran, and Amir Muzur. 2015. Historical and Social Evolution of the Healers' istorical and Social Evolution of the Healers' Charisma. *Coll. Antropol.* 39, 4 (2015), 957–963.

[97] Sarah Peters, IAN Stanley, Michael Rose, and Peter Salmon. 1998. Patients with medically unexplained symptoms: sources of patients' authority and implications for demands on medical care. *Soc. Sci. Med.* 46, 4–5 (1998), 559–565.

[98] Kathleen H Pine, Claus Bossen, Yunan Chen, Gunnar Ellingsen, Miria Grisot, Melissa Mazmanian, and Naja Holten Møller. 2018. Data Work in Healthcare: Challenges for Patients, Clinicians and Administrators. In *Companion of the 2018 ACM Conference on Computer Supported Cooperative Work and Social Computing*, 433–439.

[99] Amy S Porter, Jolene O Callaghan, Kristin A Englund, Robert R Lorenz, and Eric Kodish. 2020. Problems with the problem list: challenges of transparency in an era of patient curation. *J. Am. Med. Informatics Assoc.* 27, 6 (2020), 981–984. DOI:https://doi.org/10.1093/jamia/ocaa040

[100] John Powell. 2019. Trust me, I'm a chatbot: how artificial intelligence in health care fails the turing test. *J. Med. Internet Res.* 21, 10 (2019), e16222. DOI:https://doi.org/10.2196/16222

[101] Lucy Powley, Graham Mcilroy, Gwenda Simons, and Karim Raza. 2016. Are online symptoms checkers useful for patients with inflammatory arthritis? *BMC Musculoskelet. Disord.* 17, 1 (2016), 362. DOI:https://doi.org/10.1186/s12891-016-1189-2

[102] Samson Rajakumar and Ruth Thephila. 2019. Comparison of usability of PC based and smartphone symptom checker. Auckland University of Technology.

[103] Salman Razzaki, Adam Baker, Yura Perov, Katherine Middleton, Janie Baxter, Daniel Mullarkey, Davinder Sangar, Michael Taliercio, Mobasher Butt, and Azeem Majeed. 2018. A comparative study of artificial intelligence and human doctors for the purpose of triage and diagnosis. *arXiv Prepr. arXiv1806.10698* (2018).

[104] Soo Young Rieh. 2002. Judgment of information quality and cognitive authority in the Web. *J. Am. Soc. Inf. Sci. Technol.* 53, 2 (2002), 145–161. DOI:https://doi.org/10.1002/asi.10017

[105] Noor Van Riel, Koen Auwerx, Pieterjan Debbaut, Sanne Van Hees, and Birgitte Schoenmakers. 2017. The effect of Dr Google on doctor–patient encounters in primary care: a quantitative, observational, cross-sectional study. *BJGP open* 1, 2 (2017), 1–10. DOI:https://doi.org/10.3399/bjgpopen17X100833

[106] Lila J Finney Rutten, Neeraj K Arora, Alexis D Bakos, Noreen Aziz, and Julia Rowland. 2005. Information needs and sources of information among cancer patients: a systematic review of research (1980–2003). *Patient Educ. Couns.* 57, 3 (2005), 250–261. DOI:https://doi.org/10.1016/j.pec.2004.06.006

[107] Mark Ryan. 2020. In AI We Trust: Ethics, Artificial Intelligence, and Reliability. *Sci. Eng. Ethics* (2020), 1–19.

[108] Marcel Salathé, Thomas Wiegand, and Markus Wenzel. 2018. Focus group on artificial intelligence for health. *arXiv Prepr. arXiv1809.04797* (2018).

[109] Reijo Savolainen. 2006. Media credibility and cognitive authority: The case of seeking orienting information. *Inf. Res.* 12, 3 (2006), 12.

[110] Rebecca Schaffer, Kristine Kuczynski, and Debra Skinner. 2008. Producing genetic knowledge and citizenship through the Internet: mothers, pediatric genetics, and cybermedicine. *Sociol. Health Illn.* 30, 1 (2008), 145–159. DOI:https://doi.org/10.1111/j.1467-9566.2007.01042.x

[111] Hannah L Semigran, David M Levine, Shantanu Nundy, and Ateev Mehrotra. 2016. Comparison of Physician and Computer Diagnostic Accuracy. *JAMA Intern. Med.* 176, 12 (2016), 1860--1861. DOI:https://doi.org/10.1001/jamainternmed.2016.6001

[112] Hannah L Semigran, Jeffrey A Linder, Courtney Gidengil, and Ateev Mehrotra. 2015. Evaluation of symptom checkers for self diagnosis and triage: audit study. *bmj* 351, (2015), h3480. DOI:https://doi.org/10.1136/bmj.h3480

[113] Laura Senier, Rachael Lee, and Lauren Nicoll. 2018. The strategic defense of physician autonomy: State public health agencies as countervailing powers. *Soc. Sci. Med.* 186, (2018), 113–121. DOI:https://doi.org/10.1016/j.socscimed.2017.06.007.The

[114] Bashir Shaya, Nadine Al Homsi, Kevin Eid, Zeinab Haidar, Ali Khalil, Kelly Merheb, Gladys Honein-Abou Haidar, and Elie A Akl. 2019. Factors associated with the public's trust in physicians in the context of the Lebanese healthcare system: a qualitative study. *BMC Health Serv. Res.* 19, 1 (2019), 525.

[115] Carl Shen, Michael Nguyen, Alexander Gregor, Gloria Saza, and Anne Beattie. 2019. Accuracy of a Popular Online Symptom Checker for Ophthalmic Diagnoses. *JAMA Ophthalmol.* 137, 6 (2019), 690–692. DOI:https://doi.org/10.1001/jamaophthalmol.2019.0571

[116] Yash Raj Shrestha, Shiko M Ben-menahem, and Georg Von Krogh. 2019. Organizational Decision-Making Structures in the Age of Artificial Intelligence. *Calif. Manage. Rev.* 61, 4 (2019), 66–83. DOI:https://doi.org/10.1177/0008125619862257

[117] Keng Siau and Weiyu Wang. 2018. Building Trust in Artificial Intelligence, Machine Learning, and Robotics. *Cut. Bus. Technol. J.* 31, 2 (2018), 47–53.

[118] Souraya Sidani. 2008. Effects of patient-centered care on patient outcomes: an evaluation. *Res. Theory Nurs. Pract.* 22, 1 (2008), 24–37.

[119] C Estelle Smith, Bowen Yu, Anjali Srivastava, Aaron Halfaker, Loren Terveen, and Haiyi Zhu. 2020. Keeping Community in the Loop: Understanding Wikipedia Stakeholder Values for Machine Learning-Based Systems. In *Proceedings of the 2020 CHI Conference on Human Factors in Computing Systems*, 1–14.

[120] Felicia Wu Song, Jennifer Ellis West, Lisa Lundy, and Nicole Smith Dahmen. 2012. Women, pregnancy, and health information online: the making of informed patients and ideal mothers. *Gend. Soc.* 26, 5 (2012), 773–798. DOI:https://doi.org/10.1177/0891243212446336

[121] Tania Sourdin. 2018. Judge v. Robot: Artificial Intelligence and Judicial Decision-Making. *UNSWLJ* 41, (2018), 1114.

[122] Matt Taddy. 2018. The Technological Elements of Artificial Intelligence. *Natl. Bur. Econ. Res.* (2018).

[123] Paul C Tang and Carol Newcomb. 1998. Informing patients: a guide for providing patient health information. *J. Am. Med. Informatics Assoc.* 5, 6 (1998), 563–570.

[124] Angelique Taylor, Hee Rin Lee, Alyssa Kubota, and Laurel D Riek. 2019. Coordinating clinical teams: Using robots to empower nurses to stop the line. In *Proceedings of the ACM on Human-Computer Interaction*, 1–30.

[125] Josephine Raun Thomsen, Peter Gall Krogh, Jacob Albæk Schnedler, and Hanne Linnet. 2018. Interactive interior and proxemics thresholds: Empowering participants in sensitive conversations. In *Proceedings of the 2018 CHI Conference on Human Factors in Computing Systems*, 1–13.

[126] Tom Tomlinson, Antony J Michalski, Rebecca D Pentz, and Merja Kuuppelomäki. 2001. Futile care in oncology: when to stop trying. *Lancet Oncol.* 2, 12 (2001), 759–764.

[127] Iris D Tommelein, RE Levitt, and B Hayes-Roth. 1992. Site-layout modeling: how can artificial intelligence help? *J. Constr. Eng. Manag.* 118, 3 (1992), 594–611.

[128] Andrea Vázquez-Ingelmo, Alicia García-Holgado, Francisco José García-Peñalvo, and Roberto Therón. 2019. Dashboard meta-model for knowledge management in technological ecosystem: a case study in healthcare. In *Multidisciplinary Digital Publishing Institute Proceedings*, 44.

[129] Guy H Walker, Neville A Stanton, and Mark S Young. 2001. Where is computing driving cars? *Int. J. Hum. Comput. Interact.* 13, 2 (2001), 203–229. DOI:https://doi.org/10.1207/S15327590IJHC1302

[130] Danding Wang, Qian Yang, Ashraf Abdul, and Brian Y Lim. 2019. Designing Theory-Driven User-Centric Explainable AI. (2019), 1–15.

[131] Weiyu Wang and Keng L Siau. 2018. Living with Artificial Intelligence: Developing a Theory on Trust in Health Chatbots-Research in Progress. (2018).

[132] Mary Guptill Warren, Rose Weitz, and Stephen Kulis. 1998. Physician satisfaction in a changing health care environment: the impact of challenges to professional autonomy, authority, and dominance. *J. Health Soc. Behav.* (1998), 356–367.

[133] Elissa R Weitzman, Emily Cole, Liljana Kaci, and Kenneth D Mandl. 2011. Social but safe? Quality and safety of diabetes-related online social networks. *J. Am. Med. Informatics Assoc.* 18, 3 (2011), 292–297. DOI:https://doi.org/10.1136/jamia.2010.009712

[134] W Bradley Wendel. 2019. The Promise and Limitations of Artificial Intelligence in the Practice of Law. *Okla. L. Rev.* 72, (2019), 21.

[135] Allison Woodruff, Sarah E Fox, Steven Rousso-Schindler, and Jeffrey Warshaw. 2018. A qualitative exploration of perceptions of algorithmic fairness. In *Proceedings of the 2018 chi conference on human factors in computing systems*, 1–14.

[136] Ziang Xiao, Michelle X Zhou, Wenxi Chen, Huahai Yang, and Changyan Chi. 2020. If I Hear You Correctly: Building and Evaluating Interview Chatbots with Active Listening Skills. In *Proceedings of the 2020 CHI Conference on Human Factors in Computing Systems*, 1–14.

[137] Ji-Jiang Yang, Jianqiang Li, Jacob Mulder, Yongcai Wang, Shi Chen, Hong Wu, Qing Wang, and Hui Pan. 2015. Emerging information technologies for enhanced healthcare. *Comput. Ind.* 69, (2015), 3–11.

[138] Yue You and Xinning Gui. 2020. Self-Diagnosis through AI-enabled Chatbot-based Symptom Checkers: User Experiences and Design Considerations. In *AMIA Annual Symposium Proceedings*, American Medical Informatics Association.

[139] VHS Group Introduction. Retrieved from https://www.valurise.com/about/intro_en.html

[140] The website of Zuoshou Doctor. Retrieved from https://open.zuoshouyisheng.com/

## A  DEMOGRAPHIC INFORMATION

Table 1: Participants' demographic information.

| # | Age | Gender | AISC apps | Occupation | Consulted for whom |
|---|-----|--------|-----------|------------|--------------------|
| 1 | 29 | F | Zuoshou Doctor, WeDoctor | Logistics company staff | Self |
| 2 | 28 | F | Zuoshou Doctor, WeDoctor | Engineer | Self |
| 3 | 25 | F | Intelligent Doctor, Chunyu Doctor | MD student | Self |
| 4 | 27 | F | Zuoshou Doctor | PhD student | Self |
| 5 | 20 | M | Zuoshou Doctor，Chunyu Doctor，WeDoctor | Undergrad student | Self |
| 6 | 28 | F | Zuoshou Doctor，WeDoctor | Data analyst | Self |
| 7 | 25 | F | Zuoshou Doctor，WeDoctor | Landscape designer | Self |
| 8 | 28 | F | Zuoshou Doctor，Chunyu Doctor | Project manager | Mother, self |
| 9 | 29 | M | Zuoshou Doctor，Chunyu Doctor | Software engineer | Self |
| 10 | 28 | F | Zuoshou Doctor，Chunyu Doctor, Mighty Doctor | Product manager | Self |
| 11 | 27 | F | Zuoshou Doctor，Chunyu Doctor | Software engineer | Self |

| #  | Age | Gender | AISC apps | Occupation | Consulted for whom |
|----|-----|--------|-----------|------------|--------------------|
| 12 | 26  | F | Zuoshou Doctor，Chunyu Doctor, Mighty Doctor | Government secretary | Self |
| 13 | 28  | F | WeDoctor, Chunyu Doctor，Intelligent Doctor | Factory packer | Self |
| 14 | 28  | M | Zuoshou Doctor, WeDoctor, Mighty Doctor | Bank management trainee | Self |
| 15 | 27  | M | WeDoctor, Chunyu Doctor, Intelligent Doctor, Mighty Doctor | Software engineer | Self |
| 16 | 25  | F | Zuoshou Doctor, Intelligent Doctor | IT auditing | Self |
| 17 | 28  | M | Zuoshou Doctor，WeDoctor, Intelligent Doctor | PhD student | Self |
| 18 | 28  | F | Zuoshou Doctor, WeDoctor | Project manager | Self |
| 19 | 55  | F | Zuoshou Doctor, Intelligent Doctor | Shopkeeper | Self |
| 20 | 24  | M | Zuoshou Doctor, Chunyu Doctor | PhD student | Mother, self |
| 21 | 24  | M | Zuoshou Doctor, Chunyu Doctor | Master student | Self |
| 22 | 30  | F | Zuoshou Doctor | Freelancer | Children, self |
| 23 | 34  | M | Chunyu Doctor | Automation engineer | Children |
| 24 | 27  | M | WeDoctor, Chunyu Doctor | Salesman | Self |
| 25 | 48  | F | Zuoshou Doctor, WeDoctor | Bank clerk | Self |
| 26 | 54  | F | Zuoshou Doctor | Accountant | Self |
| 27 | 50  | F | Zuoshou Doctor | Freelancer | Spouse, friend |
| 28 | 24  | M | Zuoshou Doctor | Master of Medicine | Relative |
| 29 | 52  | F | Zuoshou Doctor | Bank clerk | Father, children, self |
| 30 | 50  | F | Zuoshou Doctor | Finance | Self |